\documentclass[twocolumn,secnumarabic,amssymb, nobibnotes, aps, prd]{revtex4}
\usepackage{graphicx}

\begin{document}
\title{Contradiction between assumption on superposition of flux-qubit states and the law of angular momentum conservation.}
\author{A.V.  Nikulov}
\affiliation{Institute of Microelectronics Technology and High Purity Materials, Russian Academy of Sciences, 142432 Chernogolovka, Moscow District, RUSSIA.} 
\begin{abstract} Superconducting loop interrupted by one or three Josephson junctions is considered in many publications as a possible quantum bit, flux qubit, which can be used for creation of quantum computer. But the assumption on superposition of two macroscopically distinct quantum states of superconducting loop contradict to the fundamental law of angular momentum conservation and the universally recognized quantum formalism. Numerous publications devoted to the flux qubit testify to an inadequate interpretation by many authors of paradoxical nature of superposition principle and the subject of quantum description.
 \end{abstract}

\maketitle

\narrowtext

\section*{Introduction}
Quantum computation and quantum information is one of the most popular themes of the last decades \cite{book2000,Nature01}. Many authors propose \cite{QubitThe01,QubitThe02} and make \cite{QubitExp01,QubitExp02,QubitExp03,Ilichev04,Mooij2003} quantum bits, main element of quantum computer, on base of different two-states quantum systems including superconducting one. The employees of D-Wave Systems Inc. claimed already that they have made the world's first commercially viable quantum computer \cite{D-Wave}. Main aim of this paper is to show that the assumptions by numerous authors on macroscopic quantum tunneling \cite{Tunnel1,Tunnel2,Tunnel3} and on superposition of two macroscopically distinct quantum states \cite{Leggett85,Leggett02,Clarke08} of superconducting loop interrupted by Josephson junctions contradict to the fundamental law of angular momentum conservation and the universally recognized quantum formalism. It is important for the problem of practical realisation of the idea of quantum computation since many authors consider such loop as possible quantum bit, flux qubit. Quantum bit is a quantum system with two permitted states, superposition of which is possible. Without  superposition of states a quantum system with two permitted states is ordinary but no quantum bit. Therefore the contradiction between the assumption on superposition and fundamental physical laws casts doubt on numerous publication about "flux qubit". These works may be unavailing. 

The contradiction between the assumption on "flux qubit" and the law of angular momentum conservation is obvious. It does not mean that this conservation law can be violated. No experimental result obtained for the present can give evidence of superposition of macroscopically distinct quantum states \cite{QI2005}. Many authors may interpret some experimental results as such evidence because of no enough profound understanding of paradoxical nature of the quantum principle of superposition \cite{QI2007}. Therefore before to consider the concrete problem of  "flux qubit" I will touch "philosophical" problems of quantum foundation and the essence of controversy between creators of quantum theory about the subject of quantum mechanics description. 

\section{What is subject of quantum mechanics description?}
For centuries science had viewed its aim as the discovery of the real. Scientists believed that they investigate an objective reality as it exists irrespective of any act of observation. But on the atomic level physicists have come into collision with paradoxical phenomena which can not be described up to now as a manifestation of an objective reality. Therefore some creators of quantum theory, Heisenberg, Bohr and others were force to advocate positivism, the point of view according to which the aim of science is investigation no objective reality but only phenomena \cite{FPP2008}. Other creators of quantum theory, Plank, Einstein, de Broglie, Schrodinger could not agree with this change of science aim. But no realistic description of quantum phenomena could be created. Therefore the quantum mechanics created at the cost of refusal of objective reality description dominates more than eighty years.  

It is important to understand that quantum mechanics, in contrast to other theories of physics, does not describe a reality. The basic principle of the idea of quantum computation was introduced in 1935 by opponents of the Copenhagen interpretation, Einstein and Schrodinger, who persisted in their opinion that the quantum theory which can describe only phenomena  can not be considered as complete. Both Einstein, Podolsky, Rosen \cite{EPR1935} and Schrodinger where sure that this principle, entanglement or Einstein - Podolsky - Rosen correlation, can not be real because of its contradiction with locality principle. Therefore Einstein, Podolsky, Rosen stated that quantum- mechanical description of physical reality can not be considered complete \cite{EPR1935} and Schrodinger introduced \cite{Schrod35} this principle as "entanglement of our knowledge" \cite{Entangl}. A "philosophical" question: "Could a real equipment be made on base of the principle which can not be describe reality?" forces to consider the essence of entanglement and history of its emergence.

\subsection{Two main paradoxes of quantum phenomena.}
Two features of quantum phenomena are most paradoxical. The both were introduced into the consideration by Einstein, the principal opponent of the Copenhagen formalism as a complete theory.

\subsubsection{Wave-particle duality}
Bohr wrote in 1949 \cite{Bohr1949}: {\it "With unfailing intuition Einstein thus was led step by step  to the conclusion that any radiation process involves the emission or absorption of individual light quanta or "photons" with energy and momentum} 
$$E = h\nu ; \ \ \ p = h\sigma \eqno{(1)}$$
{\it respectively, where $h$ is Planck's constant, while $\nu$ and $\sigma$ are the number of vibrations per unit time and the number of waves per unit length, respectively. Notwithstanding its fertility, the idea of the photon implied a quite unforeseen dilemma, since any simple corpuscular picture of radiation would obviously be irreconcilable with interference effects, which present so essential  an aspect of radiative phenomena, and which can be described only in terms of a wave picture. The acuteness of the dilemma is stressed by the fact that the interference effects offer our only means of defining the concepts of frequency and wavelength entering into the very expressions for the energy and momentum of the photon".}  

\subsubsection{Indeterminism}
Further Bohr wrote in \cite{Bohr1949}: {\it "In this situation, there could be no question of attempting a causal analysis of radiative phenomena, but only, by a combined use of the contrasting pictures, to estimate probabilities for the occurrence of the individual radiation processes".} Before \cite{Bohr1949} in 1924 \cite{Bohr1924} Bohr noted that Einstein was first who considered the individual radiation processes as spontaneous, i.e. causeless phenomenon. 

\subsection{Superposition of states as a description method of duality and of causeless phenomena.}
Advocates of the Copenhagen interpretation believe that the principle of superposition can completely describe paradoxical nature of wave-particle duality and causeless phenomena. 
\subsubsection{Double-slit interference experiment.}
Indeed, it seems that this principle can perfectly describe the duality observed in the double-slit interference experiment. If a particle with a momentum $p$ and an energy $E$ passes the double-slit as a wave $\Psi = A \exp\frac{i(pr-Et)}{\hbar}$ describing an amplitude probability then the probability 
$$P(y) = |\Psi|^{2} = |\Psi_{1}+\Psi_{2} |^{2} = A_{1}^{2} + A_{2}^{2} + 2A_{1}A_{2}\cos(\frac{pdy}{L\hbar}) \eqno{(2)}$$
to observe the arrival of the particle at a point $y$ of a detecting screen placed on a distance $L$ from two slits separated of a distance $d$ is determined by the superposition $\Psi_{1} +  \Psi_{2}$ of possibilities to pass through first $\Psi_{1} = A_{1} \exp\frac{i(pr_{1}-Et)}{\hbar}$ or second $\Psi_{2} = A_{2} \exp\frac{i(pr_{2}-Et)}{\hbar}$ slit. In accordance with this prediction all experiments give the interference patter corresponding to the momentum $p = mv$ of particles, electrons \cite{electro} with the mass $m \approx  9 \ 10^{-31} \ kg$,  neutrons \cite{neutron} with $m \approx 1.7 \ 10^{-27} \ kg$, atoms \cite{atom} with $m \approx 3.8 \ 10^{-26} \ kg$ and even massive molecules \cite{mol2003,mol2007}, for example $C_{30}H_{12}F_{30}N_{2}O_{4}$ with $m \approx 1.7 \ 10^{-24} \ kg$ and a size $a \approx 3.2 \ nm$. The interference patter appears just as a probability when particles pass one by one through the two-slit system \cite{electro}.   

\subsubsection{Probability of what?}
One may say that the wave-particle duality is observed in the double-slit interference experiment. Electron, for example, in the experiment \cite{electro} should pass the double-slit as a wave with the de Broglie wavelength $\lambda  = 1/\sigma = h/p = h/mv$ in order the interference patter of electrons distribution with a period $\Delta y \approx \lambda L/d $ can emerge at the detecting screen. But each electron is detected as particle at a point of the detecting screen. What is the essence of the de Broglie-Shcrodinger wave function $\Psi$ in this case? According to the orthodox interpretation proposed by Born $|\Psi(r,t)|^{2}$ is a probability density. But probability of what? There is possible a realistic or positivism interpretation. According to the first one $|\Psi(r,t)|^{2}dV$ is a probability that the particle is in a vicinity $dV$ of $r$ at a time $t$. According to positivism point of view such statement has no sense since quantum mechanics can describe only results of observations. The interference observations \cite{mol2007} of molecules with the size $a = 3.2 \ nm$ exceeding much its de Broglie wavelength $\lambda = h/mv \approx  0.004 \ nm$ corroborate this point of view. It is impossible to localize the molecule with the size $a \approx  3.2 \ nm$ in a volume with a size $\approx 0.1 \ nm$. We must agree with the positivism point of view that the principle of superposition can describe only results of observations and nothing besides. Therefore it is important that we have not the ghost of a chance to observe the quantum interference of a particle larger $\approx 1 \ \mu m$ \cite{FPP2008}.

\subsubsection{Radioactive decay of atom as classical example of causeless phenomena.}
Bohr wrote in \cite{Bohr1949} that "{\it in his famous article on radiative equilibrium}" published in 1917 \cite{Einstein1917} "{\it Einstein emphasized the fundamental character of the statistical description in a most suggestive way by drawing attention to the analogy between the assumptions regarding the occurrence of the spontaneous radiative transitions and the well-known laws governing transformations of radioactive substances}". Further Bohr quotes in \cite{Bohr1949} an opinion by Einstein about his theory of radiative equilibrium written at the end of the article \cite{Einstein1917}: "{\it The weakness of the theory lies in the fact that, on the one hand, no closer connection with the wave concepts is obtainable and that, on the other hand, it leaves to chance (Zufall) the time and the direction of the elementary processes}". Thus, radioactive decay of atom may be considered as classical example of causeless phenomenon the time of which is left to chance. By 1928, George Gamow had solved the theory of the alpha decay via quantum tunneling. Following Gamow, as it was made by Einstein in \cite{Einstein1949}, one can describe of uncertain state of radioactive atom with help of a superposition 
$$\Psi_{atom} = \alpha \Psi_{decay}  + \beta \Psi_{no} \eqno{(3)} $$
of decayed $\Psi_{decay}$ and not decayed $\Psi_{no}$ atom. 

\subsection{Who or what can a choice make?}
According to the positivism point of view of Heisenberg and Bohr the description of the double-slit interference experiment (2) and the radioactive decay (3) with help of the $\Psi$-function is complete. But one can agree with this point of view only if to avoid questions: "How can a particle make its way through two slits at the same time?" and "Who or what can choose result of observation?" Concerning the first question Heisenberg wrote in \cite{Heisenberg1958} {\it "A real difficulty in the understanding of the Copenhagen interpretation arises, however, when one asks the famous question: But what happens 'really' in an atomic event?"} The creators of the Copenhagen interpretation refused to answer on such question. Concerning the second question there was no agreement between they. Bohr wrote in \cite{Bohr1949} that at the Solvay meeting 1928 "{\it an interesting discussion arose also about how to speak of the appearance of phenomena for which only predictions of statistical character can be made. The question was whether, as to the occurrence of individual effects, we should adopt a terminology proposed by Dirac, that we were concerned with a choice on the part of "nature" or, as suggested by Heisenberg, we should say that we have to do with a choice on the part of the "observer" constructing the measuring instruments and reading their recording. Any such terminology would, however, appear dubious since, on the one hand, it is hardly reasonable to endow nature with volition in the ordinary sense, while, on the other hand, it is certainly not possible for the observer to influence the events which may appear under the conditions he has arranged}".  

\subsubsection{Collapse of wave function. Von Neumann's projection postulate}
The orthodox interpretation studied during last eighty years substitutes the answer on the second question with words on collapse of the wave function or a 'quantum jump' (according Heisenberg \cite{Heisenberg1958}) at observation. The necessity of the collapse postulated by von Neumann in 1932 \cite{Neumann1932} reveals the incompleteness of the Copenhagen formalism even according to the positivism point of view. The problems of wave-particle duality and indeterminism were not solved but only taken away outside the theory. The two well known paradoxes, introducing entanglement, have demonstrated clearly this incompleteness.

\subsubsection{Entanglement of two particles states in the EPR paradox demonstrates incompleteness of quantum - mechanical description of physical reality.}
A. Einstein, B. Podolsky and N. Rosen demonstrated in \cite{EPR1935} paradoxical nature of the superposition collapse using the law of conservation. In the Bohm's version \cite{Bohm1951} of the EPR paradox the spin states of two particles are entangled  
$$\Psi_{EPR} = \alpha \Psi_{\uparrow }(r_{A})\Psi_{\downarrow  }(r_{B})   + \beta \Psi_{\downarrow  }(r_{A})\Psi_{\uparrow }(r_{B}) \eqno{(4)} $$
because of the law of angular momentum conservation. Any measurement of spin projection must give opposite results independently of the distance between the particles $r_{A} - r_{B}$ since any other result means violation of this fundamental law. The description of this correlation with help of superposition and its collapse  
$$\Psi_{EPR} = \Psi_{\uparrow }(r_{A})\Psi_{\downarrow  }(r_{B})  \eqno{(5)} $$
implies that a measurement of the particle $A$ can instantly change a state of the particle $B$. This means the observation of real non-locality if superposition (4) is interpreted as description of a reality. Thus, the EPR paradox has prove unambiguously that quantum-mechanical description of physical reality can be considered complete only if non-local interaction is possible in this reality.  

\subsubsection{Entanglement of atom and cat states by Schrodinger emphasizes incompleteness of causeless phenomena description.}
In order to make obvious the incompleteness of causeless phenomena description with help of superposition (3) Schrodinger \cite{Schrod35} has entangled the states of radioactive atom and a cat with unambiguous cause - effect connection  
$$\Psi_{cat} = \alpha \Psi_{decay}G_{yes}Fl_{yes}Cat_{dead} + \beta \Psi_{no}G_{no}Fl_{no}Cat_{alive} \eqno{(6)} $$
If the atom decays $\Psi_{decay}$ then the Geiger counter tube $G_{yes}$ discharges and through a relay releases a hammer which shatter a small flask of hydrocyanic acid $Fl_{yes}$. The hydrocyanic acid should kill the cat $Cat_{dead}$. In the opposite case $\Psi_{no}$ the cat should still live $Cat_{alive}$. It is impossible logically to see that the cat is dead $Cat_{dead}$ and alive $Cat_{dead}$ at the same time. When anyone will look on the cat he should see dead 
$$\Psi_{cat} = \Psi_{decay}G_{yes}Fl_{yes}Cat_{dead} \eqno{(6a)} $$
or alive cat 
$$\Psi_{cat} = \Psi_{no}G_{no}Fl_{no}Cat_{alive} \eqno{(6b)} $$ 
The question: "Who or what can choose the cat's fate?" reveals that even causeless phenomenon must have a cause in its complete description. We must choose between {\it nature} as proposed by Dirac or {\it the observer} as suggested by Heisenberg. In the first case the description with help of superposition (6) is obviously incomplete. A natural cause because of which the atom could decay is absent the left of $\Psi_{decay}$ and $\Psi_{no}$ in (6). The suggestion of Heisenberg results to the conclusion that no reality can exist without an observer.  

\section{Can an experimental result be considered as a challenge to macroscopic realism?}
Heisenberg upheld just this absence of quantum objective reality \cite{Heisenberg1958}: {\it "In classical physics science started from the belief - or should one say from the illusion? - that we could describe the world or at least parts of the world without any reference to ourselves"}. How can one make a real equipment, which should operate without ourselves, using the quantum description, which has no sense without any reference to ourselves?  Heisenberg stated in \cite{Heisenberg1958} that {\it "there is no way of describing what happens between two consecutive observations"} and {\it "that the concept of the probability function does not allow a description of what happens between two observations"}. According to this point of view quantum mechanics can not describe the process of quantum computation which should be between observations. 

\subsection{Two different "Fathers" of quantum computing}
Thus, according to the point of view not only opponents, Einstein and Schrodinger, but also the creator of the Copenhagen formalism we have no description of the quantum computation process. Then why could this idea become so popular? The numerous publications about quantum computer result from the ideas of David Deutsch and Richard Feynman \cite{DiVincenzo}.  But it is important to note that Deutsch and Feynman have pointed different ways towards quantum computer. Deutsch invented the idea of the quantum computer in the 1970s as a way to experimentally test the "Many Universes Theory" of quantum physics - the idea that when a particle changes, it changes into all possible forms, across multiple universes \cite{Father}. This theory is one of the realistic interpretations \cite{Everett} of quantum mechanics which allows to interpreted most paradoxical quantum phenomena as manifestation of real processes. But this processes should occur across multiple universes \cite{DeutschFR}. According to Deutsch, {\it "quantum superposition is, in Many Universes terms, when an object is doing different things in different universes"} \cite{Father}. The Many Universes interpretation allows to understand why quantum computer may excel the classical one. It can do {\it "a number of computations simultaneously in different universes"} \cite{Father}. But the idea of many Universes seems mad for most physicists. Therefore most authors follow to Richard Feynman who based the idea of quantum computing on the Copenhagen interpretation. They, as well as Feynman, have an illusion, in spite of opinion of the creators, that the probability function allows a description of what happens between observations. Moreover most modern physicists are sure that quantum mechanics is an universal theory of reality from elementary particles to superconductivity. 

\subsection{What is the essence of Bell's inequality violation in?}
Einstein foresaw possibility of such mass illusion. He wrote to Schrodinger in 1928 \cite{Lett1928}: {\it "The soothing philosophy-or religion?-of Heisenberg-Bohr is so cleverly concocted that it offers the believers a soft resting pillow from which they are not easily chased away"}. Many modern authors are sure that the experimental evidence \cite{EPRexp} of violation of the Bell's inequality proves only that Einstein was not right, quantum mechanics is complete theory and we can continue to slip on the soft resting pillow proposed by Heisenberg and Bohr. But some experts understand that the experiments \cite{EPRexp} rather cast doubt on very existence of physical reality. The violation of the Bell's inequalities is sole experimental evidence of EPR correlation (entanglement) observation. In order to quantum computer could be a real equipment the entanglement must exist, but not only to be observed. But the entanglement, because of its very nature, contradict to realism, at the least local one and of single Universe.

\subsection{Doubtfulness of numerous publication about superposition and entangled states of superconductor structures.}
The absence of comprehension of these internal conflicts of the idea of quantum computer results to illusion concerning possibility to make quantum bit. Many authors are sure that it is possible not only to make qubits \cite{Mooij2003,Leggett02,Clarke08} but even to entangle their states \cite{Ilichev04}. Modern physicists have already got accustomed to the principle of superposition in the course of eighty years history of quantum mechanics in its Copenhagen interpretation. Therefore the contradiction of the assumption on superposition of macroscopically distinct quantum states with macroscopic realism \cite{Leggett85} can not trouble most of they. Many authors interpret thoughtlessly some experimental results obtained at measurements of the superconducting loop interrupted by Josephson junctions as evidence of macroscopic quantum tunneling \cite{Tunnel1,Tunnel2,Tunnel3} and superposition of states \cite{Mooij2003}. But this interpretation contradicts not only to macroscopic realism but also to fundamental law of angular momentum conservation.  

\subsubsection{Superposition of quantum states with macroscopically different angular momentum is quite impossible according to the universally recognized quantum formalism.}
Superposition and quantum tunneling are assumed between two permitted states $n$ and $n+1$ with equal energy but macroscopically different angular momentum. The angular momentum $M_{p} = (2m_{e}/e)I_{p}S$ is connected with the persistent current circulating in the loop clockwise in the $n$ permitted state and anticlockwise in the $n+1$ one \cite{Clarke08}. At the values $I_{p} \approx  5 \ 10^{-7} \ A$ and loop area $S \approx  10^{-12} \ m^{2}$ of a typical "flux qubit" \cite{Mooij2003} the angular momentum equals approximately $M_{p,n} \approx  0.5 \ 10^{5} \ \hbar $ and $M_{p,n+1} \approx  -0.5 \ 10^{5} \ \hbar $ in the $n$ and $n+1$ state. At any transition between this states the angular momentum should change on the macroscopic value $M_{p,n} - M_{p,n+1} \approx  10^{5} \ \hbar $. In spite of the obvious contradiction to the law of angular momentum conservation authors of many publications assume that this transition can be causeless, i.e. takes place through superposition of states or quantum tunneling. Such assumption can not be correct according to the universally recognized quantum formalism. 

\subsubsection{Possible assumption about an EPR pair of macroscopic systems.}
It may be that the authors of publications about "flux qubit" assume that superposition and quantum tunneling is possible thanks to a firm coupling with a large solid matrix that absorbs the change in the angular momentum, as it was made in \cite {Chudnov}. Such fantastic assumption means that states of superconducting condensate are entangled (like in the relation (4)) with a large solid matrix, i.e. the loop, substrate and so forth, of uncertainly large mass. It is impossible to take seriously such fantasy about macroscopic EPR pair.

\section*{Acknowledgement}
This work has been supported by a grant "Possible applications of new mesoscopic quantum effects for making of element basis of quantum computer, nanoelectronics and micro-system technic" of the Fundamental Research Program of ITCS department of RAS and the  Russian Foundation of Basic Research grant 08-02-99042-r-ofi.

\end{document}